\documentclass[10pt,a4paper]{article}
\usepackage{amsmath}
\usepackage{amsmath,latexsym,amssymb,amsfonts}
\usepackage[dvips]{graphicx}
\usepackage{bm}
\usepackage{xcolor}

\begin{document}

\title{\textbf{Homogeneous Instantons in Bigravity}}

\author{\textsc{Ying-li Zhang$^{a,b}$}\footnote{{\tt yingli{}@{}bao.ac.cn}},
 \textsc{Misao Sasaki$^{b}$}\footnote{{\tt misao{}@{}yukawa.kyoto-u.ac.jp}}\;\;
and \textsc{Dong-han Yeom$^{b,c}$}\footnote{{\tt innocent.yeom{}@{}gmail.com}}\\
\textit{\small{$^{a}$National Astronomy Observatories, Chinese Academy of Science,}}\\
\textit{\small{Beijing 100012, People's Republic of China}}\\
\textit{\small{$^{b}$Yukawa Institute for Theoretical Physics,
 Kyoto University, Kyoto 606-8502, Japan}}\\
\textit{\small{$^{c}$Leung Center for Cosmology and Particle Astrophysics,}}\\
\textit{\small{National Taiwan University, Taipei 10617, Taiwan}}
}

\maketitle

\begin{abstract}
We study homogeneous gravitational instantons, conventionally
called the Hawking-Moss (HM) instantons, in bigravity theory.
The HM instantons describe the amplitude of quantum tunneling from a false
vacuum to the true vacuum. Corrections to General Relativity (GR) are
found in a closed form.
Using the result, we discuss the following two issues:
reduction to the de Rham-Gabadadze-Tolley (dRGT) massive gravity and
the possibility of preference for a large $e$-folding number in the context
of the Hartle-Hawking (HH) no-boundary proposal.
In particular, concerning the dRGT limit, it is found that the
tunneling through the so-called self-accelerating branch is exponentially
suppressed relative to the normal branch, and the probability becomes zero
in the dRGT limit. As far as HM instantons are concerned,
this could imply that the reduction from bigravity to the dRGT massive gravity
is ill-defined.

\end{abstract}

\begin{flushright}
{\tt YITP-14-83}
\end{flushright}

\newpage

\tableofcontents

\newpage

\section{Introduction}

The notion of a massive spin-2 graviton mediating the gravitational
force has been the subject of much debate since the first proposal by Fierz
and Pauli~\cite{Fierz:1939}. Among many issues, there was a fatal problem
that it looked almost impossible to avoid a ghost in the scalar sector of the
theory, called the Boulware-Deser (BD)
ghost~\cite{Boulware:1972,Creminelli:2005qk,Rubakov:2008}.
A breakthrough was firstly made by a non-linear
construction of a ghost-free model by de Rham, Gabadadze
and Tolley~\cite{Rham:2010,Rham:2011PRL}, called the dRGT model,
where in the decoupling limit, the BD ghost was removed by the introduction of a Minkowski reference metric
(for a review, see~\cite{Hinterbichler:2012,deRham:2014review}).
Soon after this success, the model was generalized to the full non-linear case~\cite{Hassan:2011vm,Hassan:2012}
and the absence of the BD ghost was proved for a generic but non-dynamic reference
metric~\cite{Hassan:2011tf}. Then it was realized that a simple generalization of the non-dynamical reference metric to a dynamical one would lead to a non-linear bigravity theory without BD ghost~\cite{Hassan:2011zd,Hassan:2011pp}. After that, a
series of discoveries of the cosmological solutions and analysis of
their corresponding perturbations have been
done (see for example, ~\cite{D'Amico:2011jj}--\cite{Motohashi:2012jd} for dRGT model
and ~\cite{Volkov:2012cs}--\cite{KAAMS:2014} for bigravity).

At this stage, it is interesting to explore another cosmological
application of the theory, namely quantum transitions between different
vacua in the very early universe, particularly in the context of the cosmic
landscape~\cite{Susskind:2003kw}. It may also shed light on
the Cosmological Constant Problem (CCP) in the
landscape of vacua~\cite{Weinberg:1988cp}--\cite{PS:2012}.

Quantum transitions between vacua are described by instantons
which are solutions of the field equations with the Euclidean signature.
In the context of dRGT massive gravity, the Hawking-Moss
(HM)~\cite{Hawking:1981fz} and Coleman-De Luccia
(CDL)~\cite{Coleman:1980} instantons were studied
in~\cite{ZSS:2013,ZSYS:2013}. It was found that depending on the
choice of the model parameters, the presence of a graviton mass
may influence the tunneling rate, hence may affect the stability of a
vacuum. One of the intriguing results from the analysis of the
HM instanton is its effect on the Hartle-Hawking (HH)
no-boundary wavefunction~\cite{Hartle:1983}: in contrast to GR where
the HH no-boundary wavefunction exponentially disfavors a large
number of $e$-folds necessary for successful inflation,
the HH no-boundary wavefunction in dRGT massive gravity may have a peak at
a sufficiently large value of the Hubble parameter for which one may
obtain a sufficient number of $e$-folds of inflation~\cite{SYZ:2013}.

However, in the dRGT model, one needs to introduce a non-dynamical,
fiducial metric and fix it once and for all, which is rather unnatural.
In particular, in the context of the cosmic landscape where a variety
of geometries are probably realized, it is much more natural to
render the fiducial metric dynamical~\cite{Hassan:2011zd}.
In this paper, we investigate HM instantons, that is,
homogeneous instantons in bigravity.
We introduce two scalar fields which are minimally coupled
respectively to the physical and fiducial metrics.
We then construct a HM solution and evaluate its action.
We find that there are two branches of solutions as in the dRGT case.
For each branch we analyze the contribution from the interaction
between the physical and fiducial metrics with special attention
paid to the following two issues:

(i).
Reduction of the bigravity theory to the dRGT massive gravity.
This is the limit $M_{\mathrm{f}}^2/M_{\mathrm{P}}^2\longrightarrow\infty$
where $M_{\mathrm{f}}$ is the Planck mass associated with the gravitational
action of the fiducial metric. This limit is rather tricky because
the total Euclidean action contains the HM action of the fiducial metric
which is proportional to $M_{\mathrm{f}}^2$, which would
diverge in the dRGT limit. We find that such a divergent term
can be eliminated in one of the branches by a proper renormalization, while
it cannot be eliminated in the other branch. As discussed
in~\cite{HSS:2014ij}, this could imply that the dRGT massive gravity as the
limit of bigravity is not necessarily well-defined.

(ii).
Possible preference to a large $e$-folding number for the Hartle-Hawking (HH)
no-boundary wave function. We find that the bigravity model offers such a possibility.
The same is true in the case of dRGT gravity~\cite{SYZ:2013}, but
it seems the bigravity model has more interesting cosmological implications
because the probability depends also on the cosmological constant and
the scalar potential in the fiducial side.
Hence, a direct comparison to General Relativity (GR)
implies that in the context of bigravity, there seeems to be a much better
chance to realize the consistency between the HH no-boundary proposal and
the inflationary scenario.

It should be noted that in our model, we assume two matter
sectors coupled to the physical and fiducial metrics, respectively.
Hence, our model is free from a ghost mode, which differs from
the case where the same matter sector couples
to both metrics~\cite{YFT:2014}--\cite{GHMT:2015}.

This paper is organized as follows. In Section \ref{sec2}, we setup the
 Lagrangian for our model and formulate the equations of motion for
homogeneous (HM) instantons.
In Section \ref{sec3}, we obtain HM solutions and study their implications.
Section \ref{conclusion} is devoted to conclusion and future prospects.

Throughout the paper, the Lorentzian signature is set to be $(-, +, +, +)$.

\section{Bigravity model}\label{sec2}

We consider a bigravity model with the following action~\cite{Hassan:2011zd}:
\begin{eqnarray}\label{eq:oriaction}
S &=& \frac{1}{2} \int d^{4}x \left[ \sqrt{-g} M_{\mathrm{P}}^{2}
\left(R_{\mathrm{g}}-2\lambda_{\mathrm{g}}\right)
+\sqrt{-f} M_{\mathrm{f}}^{2}\left(R_{\mathrm{f}}-2\lambda_{\mathrm{f}}\right)
 \right] \nonumber \\
&& +~ m_{\mathrm{g}}^{2} M_{\mathrm{e}}^{2}\int d^{4}x \sqrt{-g}
\sum_{n=1}^{4}\alpha_{n} \mathcal{U}_{n}(\mathcal{K}) \nonumber \\
&& + \int d^{4}x \left[\sqrt{-g} \mathcal{L}_{\mathrm{mg}} +
\sqrt{-f} \mathcal{L}_{\mathrm{mf}} \right],
\end{eqnarray}
where $g_{\mu\nu}$ is the physical metric, $f_{\mu\nu}$ is the
fiducial metric, $M_{\mathrm{P}}$ and $M_{\mathrm{f}}$ are the Planck
masses of the physical and fiducial metrics, respectively,
and $m_{\mathrm{g}}$ is a coupling constant for the interactions
between the two metrics with $\alpha_{1}$, $\alpha_{2}$, $\alpha_{3}$
and $\alpha_{4}$ being arbitrary constants.
For the remaining quantities, $R$ is the Ricci scalar, $\lambda$
is a cosmological constant, and $\mathcal{L}_{\mathrm{m}}$ is
the matter Lagrangian, and the subscripts ${\mathrm{g}}$
and ${\mathrm{f}}$ are attached for those of the physical and fiducial
sectors, respectively. The mass $M_{\mathrm{e}}$ is defined by
\begin{align}\label{eq:Medef}
M_{\mathrm{e}}= (M_{\mathrm{P}}^{-2}+M_{\mathrm{f}}^{-2})^{-1/2}\,.
\end{align}
Thus in the dRGT massive gravity limit
$M_{\mathrm{f}}/M_{\mathrm{P}}\longrightarrow\infty$,
$m_{\mathrm{g}}$ coincides with the Fierz-Pauli mass.
As for the matter, to be specific, we focus on a minimally coupled
scalar on each side,
\begin{align}
\mathcal{L}_{\mathrm{mg}}&=
-g^{\mu\nu}\partial_{\mu}\phi_{\mathrm{g}}\partial_{\nu}\phi_{\mathrm{g}}
-V_{{\mathrm{g}}}\left(\phi_{{\mathrm{g}}}\right),\\
\mathcal{L}_{\mathrm{mf}}&=
-f^{\mu\nu}\partial_{\mu}\phi_{\mathrm{f}}\partial_{\nu}\phi_{\mathrm{f}}
-V_{{\mathrm{f}}}\left(\phi_{{\mathrm{f}}}\right).
\end{align}
The interaction terms in Eq.~(\ref{eq:oriaction}) are defined as
\footnote{We note that the action (\ref{eq:oriaction}) can be equivalently written in a more compact
way as shown in~\cite{Hassan:2011zd}.}
\begin{eqnarray}
\mathcal{U}_{1}(\mathcal{K}) &=&  \left[
\mathcal{K} \right]\equiv\mathcal{K}_\mu^\mu,\\
\mathcal{U}_{2}(\mathcal{K}) &=& \frac{1}{2!}
\left( \left[ \mathcal{K} \right]^{2} - \left[ \mathcal{K}^{2} \right] \right),
\\
\mathcal{U}_{3}(\mathcal{K}) &=& \frac{1}{3!} \left( \left[ \mathcal{K} \right]^{3}
 - 3 \left[ \mathcal{K} \right] \left[ \mathcal{K}^{2} \right]
 + 2 \left[ \mathcal{K}^{3} \right] \right),\\
\mathcal{U}_{4}(\mathcal{K}) &=& \frac{1}{4!} \left( \left[ \mathcal{K} \right]^{4}
 - 6 \left[ \mathcal{K}^{2} \right] \left[ \mathcal{K} \right]^{2}
 + 8 \left[ \mathcal{K}^{3} \right] \left[ \mathcal{K} \right]
 + 3 \left[ \mathcal{K}^{2} \right]^{2} - 6 \left[ \mathcal{K}^{4} \right] \right),
\end{eqnarray}
where $\mathcal{K}^{\mu}_{\;\;\nu} = \delta^{\mu}_{\;\;\nu} -
\left(\sqrt{g^{-1}f}\right)^\mu_\nu$.\footnote{It should be noted
that the square root expression is defined by the relationship
$\left(\sqrt{g^{-1}f}\right)^\mu_\sigma\left(\sqrt{g^{-1}f}\right)^\sigma_\nu
=g^{\mu\sigma}
f_{\sigma\nu}$.}

\subsection{Euclidean action}
By the Wick rotation $\tau = it$, the Euclidean version
of the action (\ref{eq:oriaction}) is obtained as $S_E=-iS[t=-i\tau]$.\footnote{Here we note that the `Minkowski' version
of the instanton solutions have been studied in \cite{Comelli:2011zm,Volkov:2012cs,SSEMH:2012}}
Correspondingly, in the semiclassical limit, the tunneling rate per
unit time per unit volume is expressed in terms of the Euclidean
action as
\begin{align}
    \Gamma/V= Ae^{-B}\,;
\quad
    B = S_\mathrm{E}[\bar{g}_{\mu\nu,B},\bar{\phi}_B]
-S_\mathrm{E}[\bar{g}_{\mu\nu,F},\bar{\phi}_F]\,, \label{eq:rate}
\end{align}
where $\{ \bar{g}_{\mu\nu,B},\bar{\phi}_B \}$ is the so-called
bounce solution, or an instanton, a solution of the Euclidean
equations of motion with appropriate boundary conditions, and $\{
\bar{g}_{\mu\nu,F},\bar{\phi}_F \}$ is the solution staying at the
false vacuum \cite{Coleman:1980}. Conventionally a bounce solution
$\{ \bar{g}_{\mu\nu,B},\bar{\phi}_B \}$ is explored assuming
$O(4)$-symmetry, because it is often the case that an
$O(4)$-symmetric solution gives the lowest action for a wide class
of scalar-field theories~\cite{Coleman:1977th}, hence dominates the
tunneling process. It is therefore reasonable to assume the same
even in the presence of gravity~\cite{Coleman:1980,Tanaka:1992,Lee:2009,Lee:2011}.

Here we simply extend the above assumption to the
Lorentzian-invariant bigravity theory by imposing the
$O(4)$-symmetric ansatz for both the physical and fiducial
metrics:
\begin{eqnarray}
ds^{2}_{\mathrm{g, E}} &=& N^{2}(\tau) d\tau^{2} + a^{2}(\tau) d\Omega_{3}^{2},
\label{eq:ana1}\\
ds^{2}_{\mathrm{f, E}} &=& N_{\mathrm{f}}^{2}(\tau) d\tau^{2} +
b^{2}(\tau) d\Omega_{3}^{2},\label{eq:ana2}
\end{eqnarray}
where $\tau$ is the common Euclidean time parameter for both metrics
and $d\Omega_{3}^{2}$ is the metric on a unit three sphere.

Inserting the ansatz (\ref{eq:ana1}) and (\ref{eq:ana2}) into the
Euclidean version of the action (\ref{eq:oriaction}), we obtain
\begin{align}\label{eq:action}
S_{\mathrm{E}} &= 2\pi^{2} \bigg\{ -3 M_{\mathrm{P}}^{2}\int d\tau
a\left( \frac{\dot{a}^{2}}{N} + N \right)-3 M_{\mathrm{f}}^{2}\int
d\tau b \left( \frac{\dot{b}^{2}}{N_{\mathrm{f}}}
+N_{\mathrm{f}} \right)  \nonumber \\
&  +\int d\tau a^{3}N\left[M_{\mathrm{P}}^{2}\lambda_{\mathrm{g}}+
V_{\mathrm{g}}+\frac{\dot\phi_{\mathrm{g}}^2}{2N^2}+m_g^2M_e^2\sum_{n=0}^{3}
A_{n} \left(\frac{b}{a} \right)^{n}\right] \nonumber \\
&  +\int d\tau
b^{3}N_{\mathrm{f}}\left[M_{\mathrm{f}}^{2}\lambda_{\mathrm{f}}
+V_{\mathrm{f}}+\frac{\dot\phi_{\mathrm{f}}^2}{2N_{\mathrm{f}}^2}
+m_g^2M_e^2\sum_{n=0}^{3}B_{n} \left(\frac{b}{a} \right)^{n-3}\right]\bigg\},
\end{align}
where a dot means a derivative with respect to $\tau$, and $A_{n}$
and $B_{n}$ are combinations of the parameters $\alpha_{1}$,
$\alpha_{2}$, $\alpha_{3}$ and $\alpha_{4}$ given by
\begin{eqnarray}
A_{0} &=& -4\alpha_{1}-6\alpha_{2} - 4\alpha_{3} - \alpha_{4},\label{eq:ABrel1}\\
A_{1} = 3 B_{0} &=& 3 \left(\alpha_{1}+ 3\alpha_{2} + 3\alpha_{3} + \alpha_{4}
 \right),\\
A_{2} = B_{1} &=& -3 \left( \alpha_{2} + 2\alpha_{3} + \alpha_{4} \right),\\
A_{3} = \frac{1}{3} B_{2} &=& \alpha_{3} + \alpha_{4},\\
B_{3} &=& -\alpha_{4}.\label{eq:ABrel2}
\end{eqnarray}

\subsection{Equations of motion}
By varying the action (\ref{eq:action}) with respect to $N$ and
$N_{\mathrm{f}}$, we obtain the `Friedmann' equations,
\begin{eqnarray}
\frac{\dot{a}^{2}}{N^{2}a^{2}} &=& \frac{1}{a^{2}} -
\frac{1}{3M^{2}_{\mathrm{P}}}
\left[-\frac{\dot\phi_{\mathrm{g}}^2}{2N^2}+
\Lambda_{\mathrm{g}}\left(X, \phi_{\mathrm{g}}\right) \right],
\label{eq:eqm1}\\
\frac{\dot{b}^{2}}{N_{\mathrm{f}}^{2}b^{2}} &=&
\frac{1}{b^{2}} -\frac{1}{3M^{2}_{\mathrm{f}}}
\left[-\frac{\dot\phi_{\mathrm{f}}^2}{2N_{\mathrm{f}}^2}
+\Lambda_{\mathrm{f}}\left(X,\phi_{\mathrm{f}}\right)\right],
\label{eq:eqm2}
\end{eqnarray}
where $\Lambda_{\mathrm{g}}\left(X, \phi_{\mathrm{g}}\right)$ and
$\Lambda_{\mathrm{f}}\left(X, \phi_{\mathrm{f}}\right)$ are defined as
\begin{align}
\Lambda_{\mathrm{g}}\left(X, \phi_{\mathrm{g}}\right)&\equiv
M_{\mathrm{P}}^{2}\lambda_{\mathrm{g}}^{\mathrm{eff}} +
m_g^2M_e^2\sum_{n=0}^{3} A_{n} X^{n}\,,\qquad
\lambda_{\mathrm{g}}^{\mathrm{eff}}\equiv\lambda_{\mathrm{g}}
+\frac{V_{\mathrm{g}}\left(\phi_{\mathrm{g}}\right)}{M_{\mathrm{P}}^2}\,,
\label{eq:lambdag}\\
\Lambda_{\mathrm{f}}\left(X, \phi_{\mathrm{f}}\right)&\equiv
M_{\mathrm{f}}^{2}\lambda_{\mathrm{f}}^{\mathrm{eff}} +
m_g^2M_e^2\sum_{n=0}^{3} B_{n} X^{n-3}\,,\qquad
\lambda_{\mathrm{f}}^{\mathrm{eff}}\equiv\lambda_{\mathrm{f}}
+\frac{V_{\mathrm{f}}\left(\phi_{\mathrm{f}}\right)}{M_{\mathrm{f}}^2}\,,
\label{eq:lambdaf}
\end{align}
and $X\equiv b/a$.

We note that by inserting Eqs.~(\ref{eq:eqm1}) and
(\ref{eq:eqm2}) into (\ref{eq:action}), one obtains the on-shell
action,
\begin{eqnarray}\label{eq:onshellaction}
S_{\mathrm{E}} = 4 \pi^{2} \int d\tau \bigg[aN\bigg(-3
M_{\mathrm{P}}^{2} + a^{2}\Lambda_{\mathrm{g}}\left(X,
\phi_{\mathrm{g}}\right)\bigg)+bN_{\mathrm{f}}\bigg(-3
M_{\mathrm{f}}^{2} +b^2\Lambda_{\mathrm{f}}\left(X,
\phi_{\mathrm{f}}\right)\bigg) \bigg]\,.
\end{eqnarray}
It should be noted that the interaction between physical and fiducial metrics
is encoded in $\Lambda_{\mathrm{g}}$ and $\Lambda_{\mathrm{f}}$, even though
the above action looks like the sum of two independent Einstein gravity actions.

In addition to Eqs.~(\ref{eq:eqm1}) and (\ref{eq:eqm2}), by varying with
respect to $a(\tau)$ and $b(\tau)$,
one obtains the second order differential equations,
\begin{eqnarray}
\ddot{a} &=& \frac{\dot{a}\dot{N}}{N}-
\frac{aN^2}{3M_{\mathrm{P}}^{2}}\left(\frac{\dot\phi_{\mathrm{g}}^2}{N^2}
+\Lambda_{\mathrm{g}}\right)+
\frac{m_g^2M_e^2aN}{6M_{\mathrm{P}}^{2}}
\sum_{n=0}^{3}\left(\frac{b}{a}\right)^{n}\bigg[nNA_n
+\left(n-3\right)N_{\mathrm{f}}B_n\bigg],\label{eq:tracea}\\
\ddot{b} &=& \frac{\dot{b}\dot{N_{\mathrm{f}}}}{N_{\mathrm{f}}}-
\frac{bN_{\mathrm{f}}^2}{3M_{\mathrm{f}}^{2}}
\left(\frac{\dot\phi_{\mathrm{f}}^2}{N_{\mathrm{f}}^2}+\Lambda_{\mathrm{f}}\right)
- \frac{m_g^2M_e^2bN_{\mathrm{f}}}{6M_{\mathrm{f}}^{2}}
\sum_{n=0}^{3}\left(\frac{b}{a}\right)^{n-3}
\bigg[nNA_n+\left(n-3\right)N_{\mathrm{f}}B_n\bigg]\label{eq:traceb}.
\end{eqnarray}
In the case of GR, the Friedmann equation corresponds to the Hamiltonian
constraint which represents the time reparameterization invariance.
Therefore the time derivative of it does not give a new, independent equation.
In the current case, however, only one of Eqs.~(\ref{eq:eqm1}) and (\ref{eq:eqm2})
corresponds to the Hamiltonian constraint.
Therefore, the time derivative of one of them gives a new equation which
should be consistent with the above second order differential equations.
Taking the time derivative of Eq.~(\ref{eq:eqm1}), one obtains
\begin{align}\label{eq:nfcompare}
\ddot
a&=\frac{a^2\dot\phi_{\mathrm{g}}}{6\dot{a}M_{\mathrm{P}}^2}
\left(\ddot\phi_{\mathrm{g}}+\frac{\dot{a}\dot\phi_{\mathrm{g}}}{a}
-\frac{M_{\mathrm{P}}^2\dot{\lambda}_{\mathrm{g}}^{\mathrm{eff}}}
{\dot{\phi_{\mathrm{g}}}}\right)-\frac{a\Lambda_{\mathrm{g}}}{3M_{\mathrm{P}}^2}
-\frac{m_g^2M_e^2a}{6M_{\mathrm{P}}^{2}}
\left(\frac{\dot b}{\dot a}\frac{a}{b}-1\right)
\sum_{n=0}^{3}nA_n\left(\frac{b}{a}\right)^{n}\nonumber\\
&=-\frac{a}{3M_{\mathrm{P}}^2}\left(\dot{\phi_{\mathrm{g}}}^2
+\Lambda_{\mathrm{g}}\right)-\frac{m_g^2M_e^2a}{6M_{\mathrm{P}}^{2}}
\sum_{n=0}^{3}\left(\frac{b}{a}\right)^{n}
\left[\frac{\dot b}{\dot a}(3-n)B_n-nA_n\right]\,,
\end{align}
where in the first step, we used Eq.~(\ref{eq:lambdag}) while in the
second step, Eqs.~(\ref{eq:ABrel1})--(\ref{eq:ABrel2}) are used.
Comparing Eq.~(\ref{eq:nfcompare}) with (\ref{eq:tracea}), for
consistency one finds the constraint equation,
\begin{align}
\left(\frac{\dot b}{\dot
a}-N_{\mathrm{f}}\right)\sum_{n=0}^{3}(3-n)B_n\left(\frac{b}{a}\right)^{n}=0\,.
\end{align}

The above constraint equation implies the existence of
two branches of solutions:
\begin{list}{}{}
\item[-] Branch I
    \begin{align}\label{eq:I}
        N_{\mathrm{f}} = \frac{\dot b}{\dot
a}\,,
    \end{align}
\item[-] Branch II
    \begin{align}
        &\qquad\qquad\sum_{n=0}^{3}(3-n)B_n\left(\frac{b}{a}\right)^{n}=0\,.
        %\Longrightarrow&\qquad
%\left\{ \begin{aligned}\label{eq:II}
%                  b &= X_{\pm}a\,, \\
%                          X_{\pm}&\equiv
%\frac{1+2\,\alpha_3+\alpha_4\pm\sqrt{1+\alpha_3+\alpha_3^2-\alpha_4}}
%{\alpha_3+\alpha_4}\,.
%                                  \end{aligned} \right.
\end{align}
\end{list}
In the following subsections, we discuss these two branches
separately.

\subsection{Branch I}
\label{sec2:branch1}
In this branch, the lapse function for the fiducial metric is fixed
as
\begin{eqnarray}\label{eq:b1con}
N_{\mathrm{f}} = \frac{\dot{b}}{\dot{a}}\,.
\end{eqnarray}
Combining Eqs.~(\ref{eq:eqm1}) with (\ref{eq:eqm2}) and using
Eq.~(\ref{eq:b1con}), one obtains the equation,
\begin{eqnarray}\label{eq:constrainteq}
X^2\equiv\left(\frac{b}{a}\right)^2
=\frac{M_{\mathrm{f}}^2}{M_{\mathrm{P}}^2}
\left(\frac{-\dot\phi_{\mathrm{g}}^2+2\Lambda_{\mathrm{g}}
\left(X, \phi_{\mathrm{g}}\right)
}{-\frac{\dot\phi_{\mathrm{f}}^2}{N_{\mathrm{f}}^2}+
2\Lambda_{\mathrm{f}}\left(X, \phi_{\mathrm{f}}\right)}\right)\,,
\end{eqnarray}
where we have set $N=1$ for simplicity by using the time
reparameterization invariance of the theory.
Using Eqs.~(\ref{eq:lambdag}) and (\ref{eq:lambdaf}), one can
explicitly express Eq.~(\ref{eq:constrainteq}) as an equation
containing a series of X up to 4th order in its power,
\begin{align}\label{eq:br1full}
A_3X^4&+X^3\left\{A_2-\frac{M_{\mathrm{P}}^2}{M_{\mathrm{f}}^2}
\left[B_3+\frac{1}{m_g^2M_e^2}\left(M_{\mathrm{f}}^2
\lambda_{\mathrm{f}}^{\mathrm{eff}}(\phi_{\mathrm{f}})
-\frac{\dot\phi_{\mathrm{f}}^2}{2N_{\mathrm{f}}^2}\right)\right]\right\}+X^2
\left(A_1-3A_3\frac{M_{\mathrm{P}}^2}{M_{\mathrm{f}}^2}\right)\nonumber\\
&+X\left[A_0-A_2\frac{M_{\mathrm{P}}^2}{M_{\mathrm{f}}^2}
+\frac{1}{m_g^2M_e^2}\left(M_{\mathrm{P}}^{2}
\lambda_{\mathrm{g}}^{\mathrm{eff}}(\phi_{\mathrm{g}})
-\frac{\dot\phi_{\mathrm{g}}^2}{2}\right)\right]
-\frac{A_1}{3}\frac{M_{\mathrm{P}}^2}{M_{\mathrm{f}}^2}=0\,.
\end{align}

Generally, the above equation is not easy to solve. However, in the special
case when the scalar fields $\phi_{\mathrm{g}}$ and $\phi_{\mathrm{f}}$ are
slowly varying so that we have $\dot\phi_{\mathrm{g}}^2\ll\Lambda_{\mathrm{g}}$
and $\dot\phi_{\mathrm{f}}^2/N_{\mathrm{f}}^2\ll\Lambda_{\mathrm{f}}$, we may ignore
the kinetic terms and the equation reduces to an algebra equation for $X$,
\begin{align}\label{eq:br1}
A_3X^4&+X^3\left[A_2-\frac{M_{\mathrm{P}}^2}{M_{\mathrm{f}}^2}
\left(B_3+\lambda_{\mathrm{f}}^{\mathrm{eff}}
\frac{M_{\mathrm{f}}^{2}}{m_g^2M_e^2}\right)\right]+X^2
\left(A_1-3A_3\frac{M_{\mathrm{P}}^2}{M_{\mathrm{f}}^2}\right)\nonumber\\
&+X\left(A_0-A_2\frac{M_{\mathrm{P}}^2}{M_{\mathrm{f}}^2}
+\lambda_{\mathrm{g}}^{\mathrm{eff}}\frac{M_{\mathrm{P}}^{2}}{m_g^2M_e^2}\right)
-\frac{A_1}{3}\frac{M_{\mathrm{P}}^2}{M_{\mathrm{f}}^2}=0\,.
\end{align}
Solving this equation one obtains
$X=X(\phi_{\mathrm{g}},\phi_{\mathrm{f}})=b/a$.
Thus a solution in this branch exists provided that
the above equation has a real, positive root.
In particular, in the case of our interest where $\phi_{\mathrm{g}}$
and $\phi_{\mathrm{f}}$ are homogeneous, which is the case of
our current interest, the above gives an exact solution for $X=b/a$.

Before closing this subsection, we mention a particular case of the
model parameters. As discussed in the above, Eq.~(\ref{eq:br1})
is in general an algebraic equation for $X$. However, for a particular
set of the parameters, all the coefficients of the powers of $X$ may
vanish identically. In this case $X=b/a$ becomes unconstrained.
This implies there will be substantially more varieties of solutions,
including those with non-compact or non-trivial topologies.
Interestingly, it seems this corresponds to the partially
massless bimetric theory \cite{HSS:2012pm,HSS:2012dp} where the mass
coincides with the Higuchi bound \cite{Higuchi:1987}. Furthermore,
it seems to be also related to the conformal gravity~\cite{HSS:2013pm}.
Detailed discussion on this case is beyond the scope of the present paper.
We plan to study this case in depth in a forthcoming paper~\cite{ZhYeSa}.

\subsection{Branch II}
\label{sec2:branch2}

In this branch, one obtains the algebra solution for $X$,
\begin{align}\label{eq:br2}
 \frac{b}{a}= X_{\pm}\,,
\qquad X_{\pm} \equiv
\frac{\alpha_2+2\,\alpha_3+\alpha_4\pm\sqrt{\alpha_3+\alpha_3^2
+\alpha_2\left(\alpha_3-\alpha_4\right)
-\alpha_1\left(\alpha_3+\alpha_4\right)}}{\alpha_3+\alpha_4}\,.
\end{align}
A solution in this branch exists when the model parameters are in the
range such that one of $X_\pm$ is real and positive.

We note that this branch is analogous to the `self-accelerating' branch
in the dRGT model~\cite{Gumrukcuoglu:2011open,Gumrukcuoglu:2011perturb}.
In Ref.~\cite{DeFelice:2012mx}, it was found that
this branch in dRGT massive gravity model suffers
from a ghost problem, hence is considered to be an
unhealthy branch. However, in extended massive gravity theories,
this problem may be relieved. Moreover, it is this branch which exhibits
various interesting features, including the case for the Hartle-Hawking
wave function in quantum cosmology where successful inflation may be possible
in massive gravity, in contrast to the case of GR~\cite{SYZ:2013}.
Hence we also consider the HM instantons in this branch in the following.

\section{Compact instantons: Hawking-Moss instantons}\label{sec3}
In this section, we focus on the HM instantons, that is,
compact and homogeneous instanton solutions~\cite{Hawking:1981fz}.
For the HM instantons, the scalar fields are at local maxima of their
potentials, respectively.
Therefore, $\phi_{\mathrm{g}}=\phi_{\mathrm{g,HM}}=\mathrm{constant}$,
$\phi_{\mathrm{f}}=\phi_{\mathrm{f,HM}}=\mathrm{constant}$,
$V'_{\mathrm{g}}(\phi_{\mathrm{g,HM}})=V'_{\mathrm{f}}(\phi_{\mathrm{f,HM}})=0$,
and $V''_{\mathrm{g}}(\phi_{\mathrm{g,HM}})<0$ and
$V''_{\mathrm{f}}(\phi_{\mathrm{f,HM}})<0$. From
Eqs.~(\ref{eq:eqm1}) and (\ref{eq:eqm2}), an HM solution
takes the form,
\begin{eqnarray}
a(\tau) &=& \sqrt{\frac{3M_{\mathrm{P}}^{2}}{\Lambda_{\mathrm{g,HM}}}}
\sin\left(\sqrt{\frac{\Lambda_{\mathrm{g, HM}}}{3 M_{\mathrm{P}}^{2}}}\tau\right),
\label{eq:aHM}\\
b(\tau) &=& \sqrt{\frac{3M_{\mathrm{f}}^{2}}{\Lambda_{\mathrm{f,HM}}}}
\sin\left( \sqrt{\frac{\Lambda_{\mathrm{f, HM}}}{3 M_{\mathrm{f}}^{2}}}
f(\tau)\right),
\label{eq:bHM}
\end{eqnarray}
where the function $f(\tau)$ is defined as $\dot{f}(\tau)\equiv N_{\mathrm{f}}$,
 and $\Lambda_{\mathrm{g, HM}}$ and $\Lambda_{\mathrm{f, HM}}$ are the values of
$\Lambda_{\mathrm{g}}$ and $\Lambda_{\mathrm{f}}$ at
$\phi_{\mathrm{g}}=\phi_{\mathrm{g,HM}}$ and
$\phi_{\mathrm{f}}=\phi_{\mathrm{f,HM}}$, respectively.
It has been shown in the previous section that the HM
solutions in both branches satisfy $X=b/a=\mathrm{constant}$.
Consequently, one finds the expression for $\dot f(\tau)$ as
\begin{eqnarray}\label{eq:df}
\dot f(\tau)= \frac{X\cos\left(\sqrt{\frac{\Lambda_{\mathrm{g, HM}}}
{3M_{\mathrm{P}}^{2}}}\tau\right)}{\sqrt{1-\frac{X^2M_{\mathrm{P}}^2}
{M_{\mathrm{f}}^2}\frac{\Lambda_{\mathrm{f, HM}}}
{\Lambda_{\mathrm{g, HM}}}
\sin^2\left(\sqrt{\frac{\Lambda_{\mathrm{g, HM}}}{3M_{\mathrm{P}}^{2}}}
\tau\right)}}\,.
\end{eqnarray}

On the other hand, from Eqs.~(\ref{eq:eqm1}) and (\ref{eq:eqm2}),
the time parameters $\tau$ and $f(\tau)$ can be expressed in terms
of the scale factors as
\begin{eqnarray}\label{eq:intetf}
d\tau =
\frac{da}{\sqrt{1-\frac{a^{2}\Lambda_{\mathrm{g}}(a)}{3M_{\mathrm{P}}^{2}}}}\,,
\qquad
df=
\frac{db}{\sqrt{1-\frac{b^{2}\Lambda_{\mathrm{f}}(b)}{3M_{\mathrm{f}}^{2}}}}\,.
\end{eqnarray}
Inserting these into the on-shell action~(\ref{eq:onshellaction}),
the Euclidean action may be expressed as
\begin{eqnarray}\label{eq:onshellevaluate}
S_{\mathrm{E}} = -12 \pi^{2} \left[M_{\mathrm{P}}^{2}
\int_{0}^{a_{\mathrm{max}}^{2}}
\sqrt{1-\frac{a^{2}\Lambda_{\mathrm{g}}(a)}{3M_{\mathrm{P}}^{2}}}
da^{2} +M_{\mathrm{f}}^{2} \int_{0}^{b_{\mathrm{max}}^{2}}
\sqrt{1-\frac{b^{2}\Lambda_{\mathrm{f}}(b)}{3M_{\mathrm{f}}^{2}}}
db^{2}\right],
\end{eqnarray}
where $a_{\mathrm{max}}$ and $b_{\mathrm{max}}$ are the scale
factors at their maxima.\footnote{In this equation
each of the integrals is done from 0 to its maximum, that is,
a half of the corresponding 4-sphere. Thus one should multiply it
by a factor of two to obtain the total action. This explains the
the coefficient $-12\pi^2$ instead of $-6\pi^2$.}
In the following, we compute the on-shell Euclidean action
for both branches.

\subsection{Euclidean action in Branch I}
\label{sec3:branch1}

In Branch I, from the fact that $X=b/a$ is a constant,
it is straightforward to obtain the following relation:
\begin{eqnarray}\label{eq:equalI}
N_{\mathrm{f}}=\dot{f}=\frac{\dot b}{\dot a}=\frac{b}{a}=X \,,
\end{eqnarray}
where the parameter $X$ is found to be fixed as
\begin{eqnarray}\label{eq:Xbr1}
X=\frac{M_{\mathrm{f}}}{M_{\mathrm{P}}}
\sqrt{\frac{\Lambda_{\mathrm{g, HM}}}{\Lambda_{\mathrm{f, HM}}}}
\,.
\end{eqnarray}
We note that this is consistent with Eq.~(\ref{eq:constrainteq}).
It also implies that the bubble expansion in the fiducial
metric side synchronizes with the one in the physical side,
\begin{equation} \label{eq:I-HM}
\left\{ \begin{aligned}
        a(\tau) &= \sqrt{\frac{3M_{\mathrm{P}}^{2}}{\Lambda_{\mathrm{g, HM}}}}
 \sin\left(\sqrt{\frac{\Lambda_{\mathrm{g, HM}}}{3M_{\mathrm{P}}^{2}}}~\tau\right),
 \\
        b(\tau) &= \sqrt{\frac{3M_{\mathrm{f}}^{2}}{\Lambda_{\mathrm{f, HM}}}}
\sin\left( \sqrt{\frac{\Lambda_{\mathrm{g, HM}}}{3 M_{\mathrm{P}}^{2}}}~
\tau\right)\,.
     \end{aligned} \right.
\end{equation}
Hence, the on-shell action (\ref{eq:onshellevaluate}) is obtained as
\begin{eqnarray}\label{eq:BIactionHM}
S_{\mathrm{E,~HM}}^{\mathrm{B}-\mathrm{I}}
=-24\pi^2\left(\frac{M_{\mathrm{P}}^4}{\Lambda_{\mathrm{g,
HM}}}+\frac{M_{\mathrm{f}}^4}{\Lambda_{\mathrm{f, HM}}}\right)\,,
\end{eqnarray}

From Eq.~(\ref{eq:BIactionHM}), it is obvious that in Branch I, the
system looks exactly like two copies of general relativity.
This is partly because `$b/a=\mathrm{constant}$' implies that
the interaction term between the physical and fiducial metrics
becomes a constant and it mimics an effective cosmological constant
on each side, and partly because the relation~(\ref{eq:equalI})
makes both metrics synchronize with each other, as shown in
Eq.~(\ref{eq:I-HM}).

Now let us consider the dRGT massive gravity limit,
$M_{\mathrm{f}}/M_{\mathrm{P}}\longrightarrow\infty$, in this branch.
For concreteness, we assume $\lambda_{\mathrm{f}}$ is fixed. Thus
in the limit $M_{\mathrm{f}}/M_{\mathrm{P}}\longrightarrow\infty$,
$X$ remains finite, and hence so does $M_{\mathrm{f}}/\Lambda_{\mathrm{f}}$.
Then it is obvious that the second term in Eq.~(\ref{eq:BIactionHM})
diverges in this limit. Thus one might worry if
the corresponding tunneling probability (\ref{eq:rate}) would diverge.
However, we argue that this divergence term is not a physical disaster,
but may be removed by an appropriate renormalization.

To see this, we recall Eq.~(\ref{eq:lambdaf}) where the expression for
$\Lambda_{\mathrm{f}}$ is given. If we take the limit $M_{\mathrm{f}}/M_{\mathrm{P}}\to\infty$
while keeping $M_{\mathrm{f}}^2/\Lambda_{\mathrm{f}}$ finite,
it corresponds to the limit where the all the energy scales on the fiducial
side is kept finite while the gravity there becomes infinitely heavy and
decoupled. In this limit we have
\begin{align}
\frac{M_{\mathrm{f}}^2}{\Lambda_{\mathrm{f}}}\to \frac{1}{\lambda_{\mathrm{f}}}
\quad
\mbox{as~} \frac{M_{\mathrm{f}}}{M_{\mathrm{P}}}\to\infty\,,
\end{align}
where $\lambda_{\mathrm{f}}$ is a bare cosmological constant on the
fiducial metric side. It follows that the second term in Eq.~(\ref{eq:BIactionHM})
becomes
\begin{align}
-24\pi^2\frac{M_{\mathrm{f}}^4}{\Lambda_{\mathrm{f, HM}}}
\to -24\pi\frac{M_{\mathrm{f}}^2}{\lambda_{\mathrm{f}}}\,.
\end{align}
Thus the limiting value is given solely in terms of the parameters of
the theory, namely, the gravitational and cosmological constants.
This implies one can subtract this term universally independent of
the solutions.
Namely, we define the renormalized action as
\begin{align}
S^\prime=S+24\pi\frac{M_{\mathrm{f}}^2}{\lambda_{\mathrm{f}}}\,.
\end{align}
Then for the HM solution in this branch we have
\begin{eqnarray}\label{eq:BIactionHMh}
S_{\mathrm{E,~HM}}^{\prime~\mathrm{B}-\mathrm{I}}
=-24\pi^2\left(\frac{M_{\mathrm{P}}^4}{\Lambda_{\mathrm{g,
HM}}}+\frac{M_{\mathrm{f}}^4}{\Lambda_{\mathrm{f, HM}}}
-\frac{M_{\mathrm{f}}^2}{\lambda_{\mathrm{f}}}\right)\,.
\end{eqnarray}
In the limit $M_{\mathrm{f}}/M_{\mathrm{P}}\to\infty$,
this reduces to the expression in the dRGT model,
\begin{eqnarray}\label{eq:BIdrgt}
S_{\mathrm{E,~HM}}^{\prime~\mathrm{B}-\mathrm{I}}
\longrightarrow S_{\mathrm{E,~HM,~dRGT}}^{\mathrm{B}-\mathrm{I}}
=-24\pi^2\frac{M_{\mathrm{P}}^4}{\Lambda_{\mathrm{g, HM}}}\,.
\end{eqnarray}

\subsection{Branch II}
\label{sec3:branch2}

In Branch II, using the relation $b=X_\pm a$ where $X_\pm$ given by
 Eq.~(\ref{eq:br2}), the on-shell action~(\ref{eq:onshellevaluate})
is obtained as
\begin{eqnarray}\label{eq:BIIactionHM}
S_{\mathrm{E,~HM}}^{\mathrm{B}-\mathrm{II}}=
-24\pi^2\left\{\frac{M_{\mathrm{P}}^4}{\Lambda_{\mathrm{g,\pm}}}
+\frac{M_{\mathrm{f}}^4}{\Lambda_{\mathrm{f,\pm}}}
\left[1-\left(1-\frac{X_\pm^2M_{\mathrm{P}}^2
\Lambda_{\mathrm{f,\pm}}}{M_{\mathrm{f}}^2
\Lambda_{\mathrm{g,\pm}}}\right)^\frac{3}{2}\right]\right\}\,,
\end{eqnarray}
where
\begin{align}
\Lambda_{\mathrm{g},\pm}=M_{\mathrm{P}}^{2}
\lambda_{\mathrm{g,HM}}^{\mathrm{eff}}
+m_g^2M_e^2\sum_{n=0}^{3} A_{n} X_\pm^{n}\,,\quad
\Lambda_{\mathrm{f},\pm}=M_{\mathrm{f}}^{2}
\lambda_{\mathrm{f,HM}}^{\mathrm{eff}}
+m_g^2M_e^2\sum_{n=0}^{3} B_{n} X_\pm^{n-3}\,.
\label{eq:lambdafHMb2}
\end{align}
As clear from the above expression, the correction term in this case
has a different form from that in Branch I unless the square-root term
on the right-hand side of it vanishes, which happens when
$X_\pm$ takes the value of Eq.~(\ref{eq:Xbr1}). As we have seen
in the previous subsection, this is the condition for the synchronization
of both metrics. Thus unless the two metrics are synchronized,
the form of the correction term in Branch II is different from that in Branch I.

In the massive gravity limit $M_{\mathrm{f}}/M_{\mathrm{P}}\to\infty$,
the Euclidean action reduces to
\begin{align}\label{eq:BIImassred1}
S_{\mathrm{E,~HM}}^{\mathrm{B}-\mathrm{II}}
\longrightarrow S_{\mathrm{E,~HM,~dRGT}}^{\mathrm{B}-\mathrm{II}}
-\frac{8\pi^2}{F^2}\frac{M_{\mathrm{f}}^2}{M_{\mathrm{P}}^2}
\left[1-\left(1-\alpha_{\rm HM}^2\right)^\frac{3}{2}\right]\,,
\end{align}
where
\begin{align}\label{eq:BIImassdrgt}
S_{\mathrm{E,~HM,~dRGT}}^{\mathrm{B}-\mathrm{II}}
\equiv-\frac{8\pi^2M_{\mathrm{P}}^2}{H_{\rm HM}^2}
\left[1 -\frac{X_\pm Y_{\pm}}{6}
 \left(\frac{m_g}{H_{\rm HM}}\right)^2A(\alpha_{\rm HM})\right]
\end{align}
while $F\equiv\sqrt{\lambda_{\mathrm{f}}^{\mathrm{eff}}/3}$, $H_{\rm HM}\equiv\sqrt{\Lambda_{\mathrm{g},\pm}/3M_{\mathrm{P}}^2}$,
$\alpha_{\rm HM}\equiv X_\pm F/H_{\rm HM}$ and $A(\alpha)\equiv\left[2-\sqrt{1-\alpha^2}\left(2+\alpha^2\right)\right]/\alpha^4$.
We see that the first term in Eq.~(\ref{eq:BIImassred1})
exactly coincides with the result in the dRGT massive gravity
(the detailed calculations are given in Appendix).
However, unlike the case of Branch I, the second divergent term
contains the variable $\alpha_{\rm HM}\propto H_{\rm HM}^{-1}$
which depends on the solution. This implies that it is impossible
to remove this divergence completely in the massive gravity limit.

If we remove the solution-independent universal divergence
as in the case of Branch I, we obtain
\begin{align}\label{eq:BIImassred2}
S'{}_{\mathrm{E,~HM}}^{\mathrm{B}-\mathrm{II}}
\to S_{\mathrm{E,~HM,~dRGT}}^{\mathrm{B}-\mathrm{II}}
+\frac{8\pi^2}{F^2}\frac{M_{\mathrm{f}}^2}{M_{\mathrm{P}}^2}
\left(1-\alpha_{\rm HM}^2\right)^\frac{3}{2}\,.
\end{align}
Since the second term is positive definite and divergent,
we conclude that the probability of tunneling through
the Branch II solution is exponentially suppressed and
vanishes in the dRGT massive gravity limit.
This is consistent with~\cite{HSS:2014ij} where it is
shown that this class of bigravity solutions are lost in this limit.

\subsection{Hartle-Hawking wave function}

To determine a wave function of the universe in quantum cosmology,
Hartle and Hawking proposed a boundary condition that the path integral
should be done over compact metrics with Euclidean
signature~\cite{Hwang:2013nja}. This is called the Hartle-Hawking (HH)
no-boundary proposal. However, when applied to the inflationary
universe, it predicts an exponentially small probability for a sufficiently
large number of $e$-folds which is necessary for successful inflation.

Recently we found that this deficit of the HH wave function may
be removed in dRGT massive gravity~\cite{SYZ:2013}. Namely, if
the dRGT massive gravity is realized at very high energy scales,
the correction term in the action may completely change the
behavior of the HH wave function and a sufficiently large number
of $e$-folds may be realized at high probability.
Inspired by this success in dRGT massive gravity,
in this subsection, we examine the same issue
in our bigravity model by using the HM solutions we obtained
in the previous subsections.

The HH wave function is formally given by the path integral,
\begin{eqnarray}
\Psi\left[h_{\mu\nu},\phi_{0}\right] =
\int^{\Sigma(h_{\mu\nu},\phi_0)}
\mathcal{D}g\mathcal{D}\phi\,
e^{-S_{\mathrm{E}} \left[ g_{\mu\nu}, \phi \right]},
\end{eqnarray}
where $\phi$ represents matter fields
and the integration is over all regular and compact geometries
$M$ with the boundary $\partial M=\Sigma$ on which $\partial g = h$ and
$\phi = \phi_{0}$.
This may be extended in a straightforward manner to the case of bigravity
by simply doubling the metric, $g\to (g,f)$ and
$\partial M_g\oplus\partial M_f=\Sigma(h_g,h_f,\phi_0)$ where
$(h_g,h_f)=(\partial g,\partial f)$.
Here we focus on the mini-superspace and
use the steepest-descent approximation to obtain
\begin{eqnarray}
\Psi\left[a_{0},b_{0},\phi_{0}\right]
= \int^{\Sigma(a_0,b_0,\phi_0)}
\mathcal{D}a\mathcal{D}b\mathcal{D}\phi\,
e^{-S_{\mathrm{E}} \left[ a,b, \phi \right]}
\simeq \sum_{\mathrm{sol}} e^{-S_{\mathrm{E}}\left[ a,b,\phi \right]},
\end{eqnarray}
where the sum in the last term is over on-shell solutions.
For a sufficiently flat potential the scalar field is slowly rolling,
and one may approximate the scalar field to be a constant in time,
$\phi\simeq const$ at leading order.
In this case the HH wave function depends only on the value
of $\phi$ through the effective cosmological constants for both physical and
fiducial metrics, $\Lambda_{\mathrm{g}}=\Lambda_{\mathrm{g}}(\phi)$
and $\Lambda_{\mathrm{f}}=\Lambda_{\mathrm{f}}(\phi)$.
The probability for a history that realizes $\phi=\phi_0$ is then given by
\begin{eqnarray}
P(\phi_0)
= \left|\Psi\left[\Lambda_{\mathrm{g}}(\phi_0),
 \Lambda_{\mathrm{f}}(\phi_0)\right]\right|^{2}
\propto e^{-2S_{\mathrm{E}}
\left[ \Lambda_{\mathrm{g}}(\phi_0), \Lambda_{\mathrm{f}}(\phi_0) \right]}.
\end{eqnarray}

\begin{figure}
\begin{center}
\includegraphics[scale=0.4]{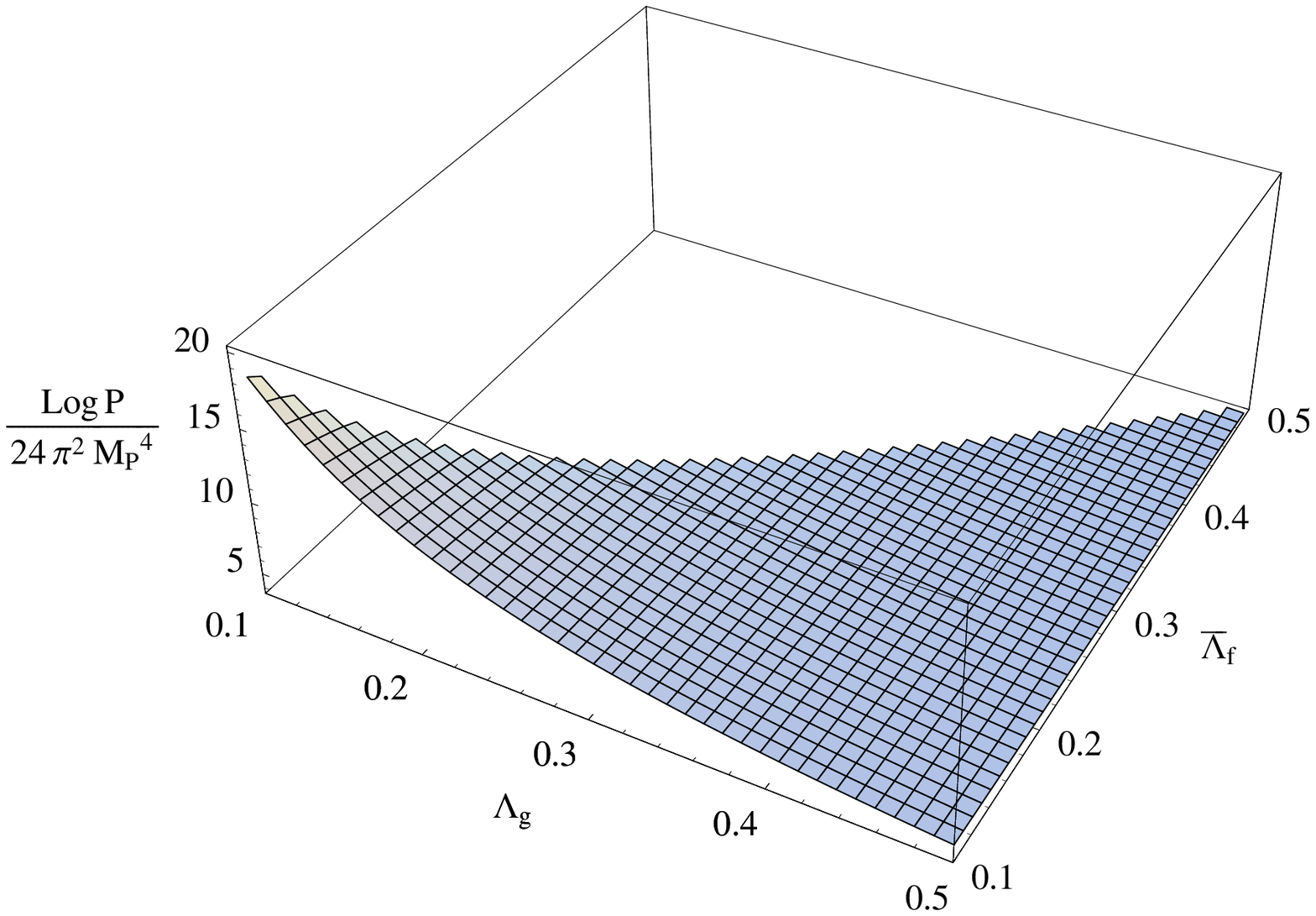}
\includegraphics[scale=0.4]{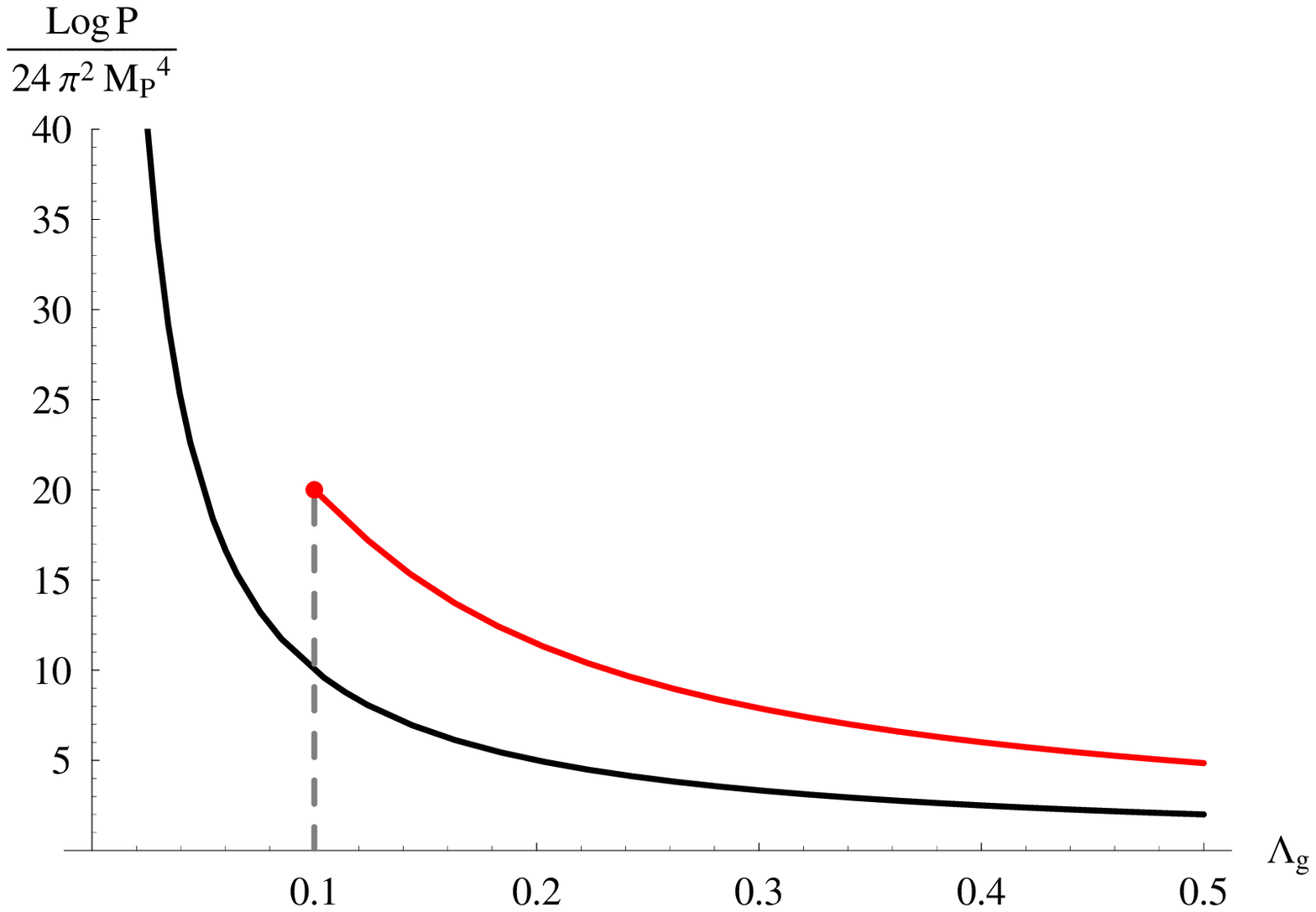}
\caption{\label{fig:probII}Left: The probability
$\log P/24 \pi^{2} M_{\mathrm{P}}^{4}$, with a fixed $\bar{X}_{\pm} = 1$.
Right: For a given fixed $\bar{\Lambda}_{\mathrm{f},\pm} = 0.1$,
the maximum probability appears around the cutoff (red),
while there is no such cutoff for Einstein gravity (black).}
\end{center}
\end{figure}

As we observed in Eq.~(\ref{eq:BIactionHM}), for Branch I,
the on-shell action is a simple sum of two independent actions,
each of which has exactly the same form as that for Einstein gravity.
Consequently the probability is dominated by the limit
$\Lambda_{\mathrm{g, HM}} \rightarrow 0$
as well as $\Lambda_{\mathrm{f, HM}} \rightarrow 0$.
On the other hand, for Branch II, the Euclidean action~(\ref{eq:BIIactionHM})
gives the probability,
\begin{eqnarray}
\frac{\log P}{24 \pi^{2} M_{\mathrm{P}}^{4}}
= \frac{1}{\Lambda_{\mathrm{g},\pm}} + \frac{1}{\bar{\Lambda}_{\mathrm{f},\pm}}
\left[ 1 - \left(1 - \bar{X}^{2}_{\pm}
\frac{\bar{\Lambda}_{\mathrm{f},\pm}}{\Lambda_{\mathrm{g},\pm}} \right)^{3/2}
\right],
\end{eqnarray}
where
\begin{eqnarray}
\bar{\Lambda}_{\mathrm{f},\pm} \equiv \Lambda_{\mathrm{f},\pm}
\frac{M_{\mathrm{P}}^{4}}{M_{\mathrm{f}}^{4}}\,,
\quad
\bar{X}_{\pm} \equiv \frac{M_{\mathrm{P}}}{M_{\mathrm{f}}} X_{\pm}\,.
\end{eqnarray}
Note that for Branch II, $X_{\pm}$ is given by the model
parameters $\alpha_{i}$, Eq.~(\ref{eq:br2}).
Hence the dependence of the probability on the solution is determined
only by the values of $\Lambda_{\mathrm{f}}$ and $\Lambda_{\mathrm{g}}$.
As we can see from the left panel of Fig.~\ref{fig:probII},
the allowed range of $\Lambda_{\mathrm{g}}$ is limited as
\begin{eqnarray}\label{eq:HHpossibility}
\Lambda_{\mathrm{g},\pm} \geq X_{\pm}^{2}
\frac{M_{\mathrm{P}}^{2}}{M_{\mathrm{f}}^{2}} \Lambda_{\mathrm{f},\pm}\,.
\end{eqnarray}
Hence, for a given $\Lambda_{\mathrm{f},\pm}$, the most probable value
becomes $\Lambda_{\mathrm{g},\pm}=X_{\pm}^{2}
\Lambda_{\mathrm{f},\pm}M_{\mathrm{P}}^{2}/M_{\mathrm{f}}^{2}$,
as shown by the right panel of Fig.~\ref{fig:probII}.
The existence of this lower cutoff of $\Lambda_{\mathrm{g}}$
for a slowly rolling scalar field indicates that
we may have an initial condition for inflation with a sufficiently large
number of $e$-folds with sufficiently high probability.

\section{Conclusion}
\label{conclusion}
As an approach to study non-perturbative effects in bigravity,
we considered quantum tunneling by introducing two tunneling fields,
respectively, minimally coupled to the physical and fiducial metrics.
Then we derived for the Hawking-Moss (HM) instanton solutions.
For a fixed set of the model parameters, we found two branches of solutions.
We called these two branches as Branch I and II, respectively,
and discussed their properties and implications.

First, we considered the dRGT massive gravity limit,
 $M_{\mathrm{f}}/M_{\mathrm{P}}\longrightarrow\infty$,
where the fiducial metric becomes non-dynamical.
In this limit we found that the action diverges as $\propto M_{\mathrm{f}}^2$,
but in Branch I, the divergent term can be eliminated by a proper renormalization
and the corresponding result in dRGT gravity is smoothly recovered.
However, in Branch II, we found that the divergent term cannot be renormalized.
Namely, there exists a solution-dependent divergence in the Euclidean action.
This branch corresponds to the self-accelerating branch in the dRGT limit.
Since this divergence is found to be positive definite,
it implies that the probability of finding this branch is exponentially
suppressed as we approach dRGT massive gravity. In dRGT massive gravity,
the self-accelerating branch is known to be unstable due to the existence
of a ghost mode \cite{DeFelice:2012mx}. Our result is quite interesting in this respect.
It suggests that the self-accelerating branch may be avoided quantum
cosmologically, if dRGT massive gravity is regarded as a limit in bigravity.
On the other hand, this could also imply that
the reduction from bigravity to the dRGT massive gravity may not be well-defined.

Second, as a direct application of the HM solution, we considered
the wave function of the universe with the Hartle-Hawking (HH) no-boundary
boundary condition. In this case, there is essentially no difference in
the prediction of the Branch I solution from that of Einstein gravity.
Namely, the HH wave function predicts the number of $e$-folds which is
too small to make inflation successful.
On the other hand, for the Branch II solution we found that
the probability of realizing a sufficiently large number of $e$-folds
becomes non-negligible, at least not exponentially suppressed.
This suggests that the HH no boundary proposal may be saved in the context
of bigravity.

It would be natural to go further to investigate the Coleman-De Luccia
instantons in bigravity theory. In this case, the matter field is no
more homogeneous and hence Eq.~(\ref{eq:br1full}) is no more an algebraic
equation. That is, $b/a$ is no longer a constant but varies as the scalar
field varies. This makes the problem much more difficult to solve.
We would like to come back to this topic in future.

Finally, as we mentioned at the end of Sec.~\ref{sec2:branch1},
for a particular case of the model parameters,
$X=b/a$ becomes unconstrained. Hence this case will allow
a lot more varieties of solutions, and may have intriguing
cosmological implications \cite{HSS:2014ij}.
Detailed discussion on this case is given in a forthcoming paper~\cite{ZhYeSa}.

\section*{Acknowledgment}
We would like to thank Antonio De Felice, Kazuya Koyama, Shinji Mukohyama,
Ryo Saito, Takahiro Tanaka and Gong-bo Zhao for valuable discussion.
We are also grateful to an anonimous referee for useful comments.
This work was supported by the
JSPS Grant-in-Aid for Scientific Research (A) No. 21244033.
DY and YZ would like to thank Bum-Hoon Lee and Wonwoo Lee for supporting to visit
Center for Quantum Spacetime, Sogang University.
DY is also supported by Leung Center for Cosmology and Particle Astrophysics
(LeCosPA) of National Taiwan University (103R4000). YZ is also supported by the Strategic Priority Research Program
``The Emergence of Cosmological Structures" of the Chinese
Academy of Sciences, Grant No. XDB09000000.

%%%%%%%%%%%%%%%
\section*{Appendix: Massive gravity limit of Branch II}\label{HMredbranch2}
%%%%%%%%%%%%%%%
In this Appendix, we derive the dRGT massive gravity limit of
the HM solutions given in Eq.~(\ref{eq:BIIactionHM}).
For convenience, here we set $\alpha_1=0$ and $\alpha_1=1$
so that the value for $X_\pm$ coincides with that in the dRGT model.
Since the first term in the curly brackets is nothing but the one for
the conventional GR case, we focus on the reduction of
the second term in the following.

In the limit $s\equiv M_{\mathrm{P}}^2/M_{\mathrm{f}}^2\ll1$, recalling
that $M_{\mathrm{e}}\longrightarrow M_{\mathrm{P}}$ from Eq.~(\ref{eq:Medef}),
we have
\begin{align}\label{eq:BIImass}
\frac{M_{\mathrm{f}}^4}{\Lambda_{\mathrm{f,\pm}}}
&\left[1-\left(1-\frac{X_\pm^2M_{\mathrm{P}}^2\Lambda_{\mathrm{f,\pm}}}
{M_{\mathrm{f}}^2\Lambda_{\mathrm{g,\pm}}}\right)^\frac{3}{2}\right]
\nonumber\\
&=\frac{M_{\mathrm{P}}^2}{s\left(\lambda_{\mathrm{f, HM}}^{\mathrm{eff}}
+\frac{m_{\mathrm{g}}^2Y_\pm}{X^3}s\right)}
\left\{1-\left[1-\frac{X_\pm^2M_{\mathrm{P}}^2}
{\Lambda_{\mathrm{g,\pm}}}\left(\lambda_{\mathrm{f, HM}}^{\mathrm{eff}}
+\frac{m_{\mathrm{g}}^2Y_\pm}{X^3}s\right)\right]^\frac{3}{2}\right\}
\nonumber\\
&=\frac{M_{\mathrm{P}}^2}{s\lambda_{\mathrm{f, HM}}^{\mathrm{eff}}}
\left(1-\frac{m_{\mathrm{g}}^2Y}{X^3\lambda_{\mathrm{f, HM}}^{\mathrm{eff}}}s
+\mathcal{O}\left(s^2\right)\right)\left[
1-\gamma^{\frac{3}{2}}\left(1-\frac{3\beta}{2} s\right)
+\mathcal{O}\left(s^2\right)\right]\nonumber\\
&=\frac{M_{\mathrm{P}}^2}{s\lambda_{\mathrm{f, HM}}^{\mathrm{eff}}}
\left[1-\gamma^{\frac{3}{2}}+s\left(\frac{3}{2}\gamma^{\frac{3}{2}}\beta
+\frac{m_{\mathrm{g}}^2Y_\pm}{X^3\lambda_{\mathrm{f, HM}}^{\mathrm{eff}}}
\left(\gamma^{\frac{3}{2}}-1\right)\right)+\mathcal{O}\left(s^2\right)\right]
\,,
\end{align}
where, for notational simplicity, we have introduced
\begin{align}\label{eq:syb}
\beta&\equiv\frac{m_{\mathrm{g}}^2
M_{\mathrm{P}}^2Y_\pm}{\gamma X_\pm\Lambda_{\mathrm{g,\pm}}}\,,\quad
\gamma\equiv1-\frac{M_{\mathrm{P}}^2
X_\pm^2\lambda_{\mathrm{f, HM}}^{\mathrm{eff}}}{\Lambda_{\mathrm{g,\pm}}}\,,
\\
Y_\pm&\equiv3(1-X_{\pm})+3\alpha_3(1-X_{\pm})^2+\alpha_4(1-X_{\pm})^3
=X_{\pm}^3\sum_{n=0}^{3} B_{n} X_{\pm}^{n-3}\,.
\end{align}

In order to compare the above with the result of dRGT massive gravity, we
first note that from Eq.~(\ref{eq:bHM}), the scale factor $b(\tau)$ reduces as
\begin{eqnarray}
b(\tau)\longrightarrow\sqrt{\frac{3}{\lambda_{\mathrm{f}}^{\mathrm{eff}}}}
\sin\left( \sqrt{\frac{\lambda_{\mathrm{f}}^{\mathrm{eff}}}{3}}
f(\tau)\right)\,,
\label{eq:bmass}
\end{eqnarray}
while in dRGT massive gravity we have
$b=F^{-1}\sin(Ff)$ with the fiducial Hubble parameter $F$.
It is obvious that $\sqrt{\lambda_{\mathrm{f}}^{\mathrm{eff}}/3}$ plays
the role of $F$,\footnote{It should be noted that in Ref.~\cite{ZSS:2013},
the Lorentzian signature of the fiducial metric is kept as it is throughout
the computation since the fiducial metric is non-dynamical in dRGT massive gravity.
However, it is dynamical in bigravity, so we should Wick rotate it
as has been done in Eq.~(\ref{eq:ana2}).
This introduces the appearance of an imaginary number in the function
$f(\tau)$ which makes the hyperbolic function transform into
the trigonometric function $b=F^{-1}\sin(Ff)$.}
\begin{eqnarray}\label{eq:lfred}
\lambda_{\mathrm{f}}^{\mathrm{eff}}\longleftrightarrow3F^2,
\end{eqnarray}
while $\Lambda_{\mathrm{g,\pm}}$ corresponds to the HM Hubble
parameter,
\begin{eqnarray}\label{eq:lgred}
\Lambda_{\mathrm{g,\pm}}\longleftrightarrow3M_{\mathrm{P}}^2H_{\mathrm{HM}}^2.
\end{eqnarray}
Hence, using Eqs.~(\ref{eq:lfred}) and (\ref{eq:lgred}), and from the
definition of the dimensionless parameter
$\alpha\equiv X_\pm F/H$, one finds the correspondences,
\begin{eqnarray}\label{eq:Xred}
\frac{M_{\mathrm{P}}^2X_\pm^2
\lambda_{\mathrm{f}}^{\mathrm{eff}}}{\Lambda_{\mathrm{g,\pm}}}
\longleftrightarrow\alpha^2
\quad\Longrightarrow\quad\gamma\longleftrightarrow1-\alpha^2\,.
\end{eqnarray}

Inserting Eqs.~(\ref{eq:lfred})--(\ref{eq:Xred}) into (\ref{eq:BIImass}),
one finds
\begin{align}\label{eq:BIImassred}
\frac{M_{\mathrm{f}}^4}{\Lambda_{\mathrm{f,\pm}}}
&\left[1-\left(1-\frac{X_\pm^2M_{\mathrm{P}}^2\Lambda_{\mathrm{f,\pm}}}
{M_{\mathrm{f}}^2\Lambda_{\mathrm{g,\pm}}}\right)^\frac{3}{2}\right]
\nonumber\\
&\longrightarrow\frac{M_{\mathrm{P}}^2}{3F^2s}
\left[1-\left(1-\alpha^2\right)^\frac{3}{2}\right]
+\frac{m_g^2M_{\mathrm{P}}^2X_\pm Y_\pm}
{18H^4_{\mathrm{HM}}\alpha^4}\left[\sqrt{1-\alpha^2}(2+\alpha^2)-2\right]\,.
\end{align}
Inserting the above into Eq.~(\ref{eq:BIIactionHM}),
one finally obtains the HM action in the dRGT massive gravity limit,
\begin{align}\label{eq:BIIactionred}
&S_{\mathrm{E,~HM}}^{\mathrm{B}-\mathrm{II}}
\nonumber\\
&\quad\longrightarrow-\frac{8\pi^2M_{\mathrm{P}}^2}{H_{\rm HM}^2}
\left[1 -\frac{X_\pm Y_{\pm}}{6}
\left(\frac{m_g}{H_{\rm HM}}\right)^2A(\alpha_{\rm HM})\right]
-\frac{8\pi^2}{F^2}\frac{M_{\mathrm{f}}^2}{M_{\mathrm{P}}^2}
\left[1-\left(1-\alpha_{\rm HM}^2\right)^\frac{3}{2}\right]\,,
\end{align}
where the function $A(\alpha)$ is defined by
\begin{align}
A(\alpha)\equiv\frac{\left[2-\sqrt{1-\alpha^2}(2+\alpha^2)\right]}{\alpha^4}\,.
\end{align}
Comparing Eq.~(\ref{eq:BIIactionred}) with Eq.~(4.14) of Ref.~\cite{ZSS:2013},
it is obvious that the first square brackets agrees with
the result in dRGT massive gravity. However, unlike the case in Branch I, the
second divergent term cannot be eliminated universally
since it contains the solution-dependent variable
$\alpha_{\rm HM}\propto H_{\rm HM}^{-1}$.

\end{document}